\documentclass{ieee}

\usepackage{times}
\usepackage{epsfig}
\usepackage{amsmath}
\usepackage{amsfonts}
\usepackage{graphics}
\usepackage{times}
\usepackage{algorithm}
\usepackage{algorithmic}
\usepackage{subfigure}
\usepackage{alltt}
\usepackage{amssymb}
\usepackage{latexsym}
\usepackage{fancyvrb}
\usepackage[small, hang]{caption2}
\usepackage{tabularx, array}
\usepackage{float}
\usepackage{natbib}
\usepackage{sectsty}
\usepackage{titletoc, minitoc}
\usepackage{longtable}
\usepackage{lastpage}

\newcommand{\ebnf}[2]{\noindent\begin{scriptsize}
    \begin{alltt}
      \begin{list}{}{
          \setlength{\rightmargin}{0cm}
          \setlength{\topsep}{0cm}
          \settowidth{\labelwidth}{\tt #1}
          \setlength{\labelsep}{0cm}}
      \item #1 $\rightarrow$ #2
    \end{list}
  \end{alltt}
  \end{scriptsize}
}

\DefineVerbatimEnvironment{code}{Verbatim}{fontsize=\footnotesize}
\graphicspath{{images/}}

\begin{document}

\title{Enabling Agents to Dynamically Select Protocols for Interactions}

\author{
  Jos\'{e}~Ghislain~Quenum, Samir~Aknine\\
  Laboratoire d'Informatique de Paris6\\ 75015 Paris, France\\
  \texttt{\{jose.quenum, samir.aknine\}@lip6.fr}\\
}


\maketitle
\thispagestyle{empty}

\begin{abstract}
  
  The work achieved in multi-agent interactions design mostly relates
  to protocols definition, specification, etc.  In this paper we
  tackle a new problem, \emph{the dynamic selection of interaction
    protocols}.  Generally the protocols and the roles agents play in
  protocol based interactions are imposed upon the system at design
  time.  This selection mode severely limits the system's openness,
  the dynamic behaviours agents can exhibit, the integration of new
  protocols, etc.  To address this issue, we developed a method which
  enables agents to select protocols themselves at runtime when they
  need to interact with one another.
  Regarding the conditions which hold in the MAS 
  agents can either \emph{jointly} perform the selection or
  \emph{individually}. We define the concepts and algorithms which
  enable agents to perform this dynamic selection. We also describe
  the mechanisms which help agents anticipate interaction
  inconsistencies.

\end{abstract}

\section{Introduction}

Generally, the interaction protocols which support agents
collaborative tasks' execution are imposed upon multi-agent systems
(MAS) at design time.
This \emph{static} protocols selection severely limits the system's
openness, the dynamic behaviours agents can exhibit, the integration
of new protocols, etc. For example, consider a collaborative task
which can be executed following varied methods either by means of a
Request protocol~\cite{FIPA:01b} (an identified agent exhibiting
specific skills is requested to perform the task) or a Contract Net
protocol (CNP)~\cite{Smith:80} (a competition holds between some
identified agents in order to find out the best one to perform the
task). Since the MAS is open and CNP looks for the best contractor, it
will undoubtedly be preferred to Request for such a task in absence of
constraints such as execution delay.  Thus, selecting Request at
design-time prevents the initiator agent from benefiting a better
processing of this task.  Moreover, the set of protocols used in
multi-agent interactions is increasingly enlarging, and a static
protocol selection will only consider the protocols the designer knows
even if these agents have the capacity of interpreting and executing
other protocols unknown at design time.  To overcome this limitation,
we should enable agents to dynamically select protocols in order to
interact.

As yet, there have been some efforts~\cite{Bourne:00,Durfee:99} to
enable agents to dynamically select the roles they play during
interactions using Markov Decision Processes, planning or even
probabilistic approaches.  However, they don't suit protocol based
coordination mechanisms. Indeed, as protocols are partially sorted
sequences of pre-formatted messages exchange, selecting them to
execute a task requires that their descriptions match that of the
collaborative task.  The solutions proposed so far do not explicitly
focus on protocols and do not check such compliance either. To address
this void, we developed a method which enables agents to dynamically
select protocols and roles in order to interact. Our method puts the
usual assumptions about multi-agent interactions a step further.
First, we consider that some interaction protocols can be known only
at runtime. Thus, starting from a minimal version, agents interaction
models can grow up by integrating these protocols from safe and
authenticated libraries of interaction protocols when needed.  Second,
we consider that agents may have different designers, therefore they
may encompass different protocols specification formalisms.
Furthermore, an interaction protocol is a triple $\{\mathbf{R},
\mathbf{M}, \mathbf{\Omega}\}$ where $\mathbf{R}$ is a set of
interacting roles which can be of two types: \emph{initiator} and
\emph{participant}. An \emph{initiator} role is the unique role in
charge of starting the protocol whereas a \emph{participant} role is
any role taking part in the protocol. Consequently, an \emph{initiator
  agent} will be any agent playing the initiator role in an
interaction while a \emph{participant agent} will be any agent taking
up a participant role. In addition, protocols used in MAS can be
classified\footnote{A complex protocol can be a combination of these
  basic categories.} in three categories: (1) \emph{1-1 protocols},
which are protocols made of two roles (initiator and participant) both
of them having only one instance (ex Request); (2) \emph{1-1$^N$
  protocols}, again protocols made of two roles with several instances
of the participant (ex CNP); (3) \emph{1-N protocols}, which are
protocols with several distinct participant roles each of them having
only one instance (ex an auction protocol with one buyer, one seller
and one manager).  Subsequently, we define the dynamic protocols
selection process for each of these categories.

Rather than explicitly indicating the protocols and the roles to use
for all the agents which will execute the desired interaction, we
suggest that agents programmers simply mention the collaborative
task's description in the \emph{initiator agent}'s source code.  As
soon as an \emph{agent} locates such a description, it identifies some
\emph{potential participant agents} and thereafter fires the dynamic
protocol selection process taking up the initiator role.  We assume
that the MAS is provided with potential participant agents
identification procedures.  
Agents can dynamically select protocols in two possible ways.  First,
the \emph{initiator agent} and all the potential \emph{participant
  agents} collectively select a protocol and assign roles to each
agent inside this protocol. This is the \emph{joint protocol
  selection} method which assumes that agents trust one another and
that they don't dread publishing their knowledge and preferences.  On
the other hand, agents can individually select protocols and roles and
start the desired interaction. This is the \emph{individual protocol
  selection} method which assumes that agents do not trust one another
and/or the system is heterogeneous (several sub systems with different
protocol formalisms are plugged together). In this method, as the
selected roles may mismatch, agents should anticipate errors in order
to guarantee consistent messages exchange. We focus on \emph{wrong
  message structure error} which indicates that something is wrong in
the message structure (performative, content, language, ontology,
etc.) and \emph{wrong message content error} which indicates that the
message's content doesn't match the expected content pattern. We argue
that our method introduces more flexibility in protocols execution,
fosters agents autonomy, favours their dynamic behaviours and suits
MAS' openness. In this paper we describe both methods and detail their
principles, concepts and algorithms.  We exemplify them towards a web
documents filtering MAS composed of (1) \emph{query agents}
representing the queries users formulate, (2) \emph{document agents}
representing the documents retrieved from the web, (3) and \emph{rule
  agents} corresponding to any linguistic rule invoked to compute
documents attributes ({\tt author(s)}, {\tt
  content}, {\tt language}, etc.).  


The paper is organised as follows. Section~\ref{sect:probdesc}
formally defines the protocols selection problem.
Sections~\ref{sect:dynselj}~and~\ref{sect:dynseli} detail our methods.
Section~\ref{sec:related} discusses some related work and
section~\ref{sec:conclusion} draws some conclusions.

\label{sect:probdesc}

\section{Problem Description}
\label{sect:probdesc}

Our purpose in this research, is to ease protocols definition,
implementation and use.  Thus, we claim to free agents programmers
from hard coding the protocols and the exact roles to use every time
their agents have to execute a collaborative task. Rather, they should
only mention in the initiator agent's source code the description of
the collaborative task to execute.  Then, once an agent comes across
such a description, it will launch the dynamic selection process
implicating potential participant agents. Concretely, given a
collaborative task $t_j$ which is to be executed by a set
$\mathcal{A}=\{a_1, a_2, \ldots a_k\}$ of agents, the selection
problem is stated as \emph{ how an agent can select a protocol and a
  role inside this protocol to execute $t_j$}? It consists in finding
out a protocol $p$ and the roles $r_1, r_2, \ldots r_p$ each $a_i$
should be enacting in this protocol in order to get $t_j$ executed. We
assume that each agent is provided with an \emph{interaction model}
$\mathcal{I}=\{r_1, r_2, \ldots r_n\}$ containing some configured
protocols ($p_1 \ldots p_m$) some of which can be introduced at
runtime and for each protocol $p_i$ a set of roles $\{r_1, r_2, \ldots
r_k\}$ this agent can play during interactions based on $p_i$.

In the following two sections, we elaborate on our solution to the
selection problem.

\section{Joint Protocol Selection}
\label{sect:dynselj}

Once an agent locates the description of a collaborative task, it
finds out a set of protocols needed to execute this task which it
refines to a sub set of protocols whose initiator roles are configured
inside its interaction model. Moving from collaborative tasks models
to protocols' requires the agents to analyse both models and detect
their adequacy.  In this paper we assume that agents are able to
examine tasks and protocols models and relate the first ones to the
second ones. After moving from task to protocols the initiator agent
should identify all the potential participant agents for the
determined protocols.  Both steps provide the initiator agent with a
sparse matrix: potential participants linked to protocols.  These
potential participant agents are thus contacted whether at the same
time or one after the
other and are required to validate a protocol. 
To contrast the messages exchanged during the joint selection and
those exchanged during normal interactions, we proposed some
performatives which we informally describe here bellow:

\begin{description}
  
\item[{\tt call-for-collaboration}]the sender of this performative
  invites the receiver to take part in a protocol described in the
  {\tt content} field.
  
\item[{\tt unable-to-select}]the sender of this performative informs
  the receiver that it cannot play a participant role in the related
  protocol. The {\tt in-reply-to} and {\tt reply-with} fields help
  relate this message to a prior {\tt call-for-collaboration}. The
  reasons why an agent may reply this performative, though identified
  as a potential participant for the protocol, are (1) its autonomy
  since it may not want to execute this protocol at this moment and
  (2) some errors in some fields.
  
\item[{\tt stop-selection}]the sender of this performative asks the
  receiver to stop the selection process this message is linked to.
  
\item[{\tt ready-to-select}]the sender of this performative notifies
  the receiver of the participant roles it can be enacting regarding
  the protocol description it received. All the participant roles in
  any protocol compatible with the current one can be listed. This
  grouping not only reveals by order of preference the roles the
  sender commits in playing but it also avoid going back and forth
  about protocols sharing the same background. Roles of protocols are
  compatible when they can execute safe interactions albeit the
  difference in their respective specifications.  As an example the
  initiator role of CNP can interact with either the participant role
  of CNP or that of Iterated CNP (ICNP~\cite{FIPA:01b}). While the
  initiator of ICNP can't interact with the participant of CNP because
  of the probable iterations.
  
\item[{\tt notify-assignment}]the sender of this performative informs
  the receiver about the role the latter has been assigned to in the
  jointly selected protocol. The assigned role is one among those the
  receiver priorly committed in playing.

\end{description}

Whatever protocol category the selection is concerned with, we can
describe the joint selection messages exchange sequence as follows:

\begin{enumerate}
  
\item the initiator agent sends a {\tt call-for-collaboration}
  encapsulating a protocol's description.
  
\item Each participant agent can reply with an {\tt unable-to-select}
  driving the initiator agent to stop the selection process between
  both agents by sending a {\tt stop-selection}.
  
\item Each participant agent can also reply with a {\tt
    ready-to-select}. In this case, the initiator agent parses the
  participant's proposals and adopts one of them sending a {\tt
    notify-assignment} or reject all the proposals sending a {\tt
    stop-selection}.

\end{enumerate}

In the remainder of this section we detail the joint selection method
for each class of protocol.

\subsection{Inside the \emph{Joint Protocol Selection}}

\subsubsection{\emph{1-1} Protocols}
\label{sect:jsel11}

In the dynamic \emph{1-1 protocols} selection, the aim of the
initiator agent is to early find out a solution, a couple $(a_i, p_j)$
where $a_i$ is one of the potential participant agents formerly
identified and $p_j$ one of the \emph{1-1 protocols} determined for
the current task. In the midst of the solution search is the matrix's
exploration. Hence, it behoves the initiator agent to explore the
matrix traversing protocols or potential participant agents. In the
protocol-oriented exploration, the initiator selects a protocol and
iterates through the set of agents which it identified for this
protocol and retains one that fits the protocol.  As soon as an agent
is determined the selection process successfully completes.
Otherwise, the iteration proceeds until there is no more protocol to
select.  Analogously, in the agent-oriented exploration the initiator
selects an agent and delves into its protocols' set looking for one
they can execute together. Whatever exploration the initiator adopts,
it should overcome the matrix sparsity by selecting as next element
(protocol or agent) the one holding the least sparse vector.

\begin{figure}[hhhh] 
\centering \includegraphics[width=.2\textwidth]{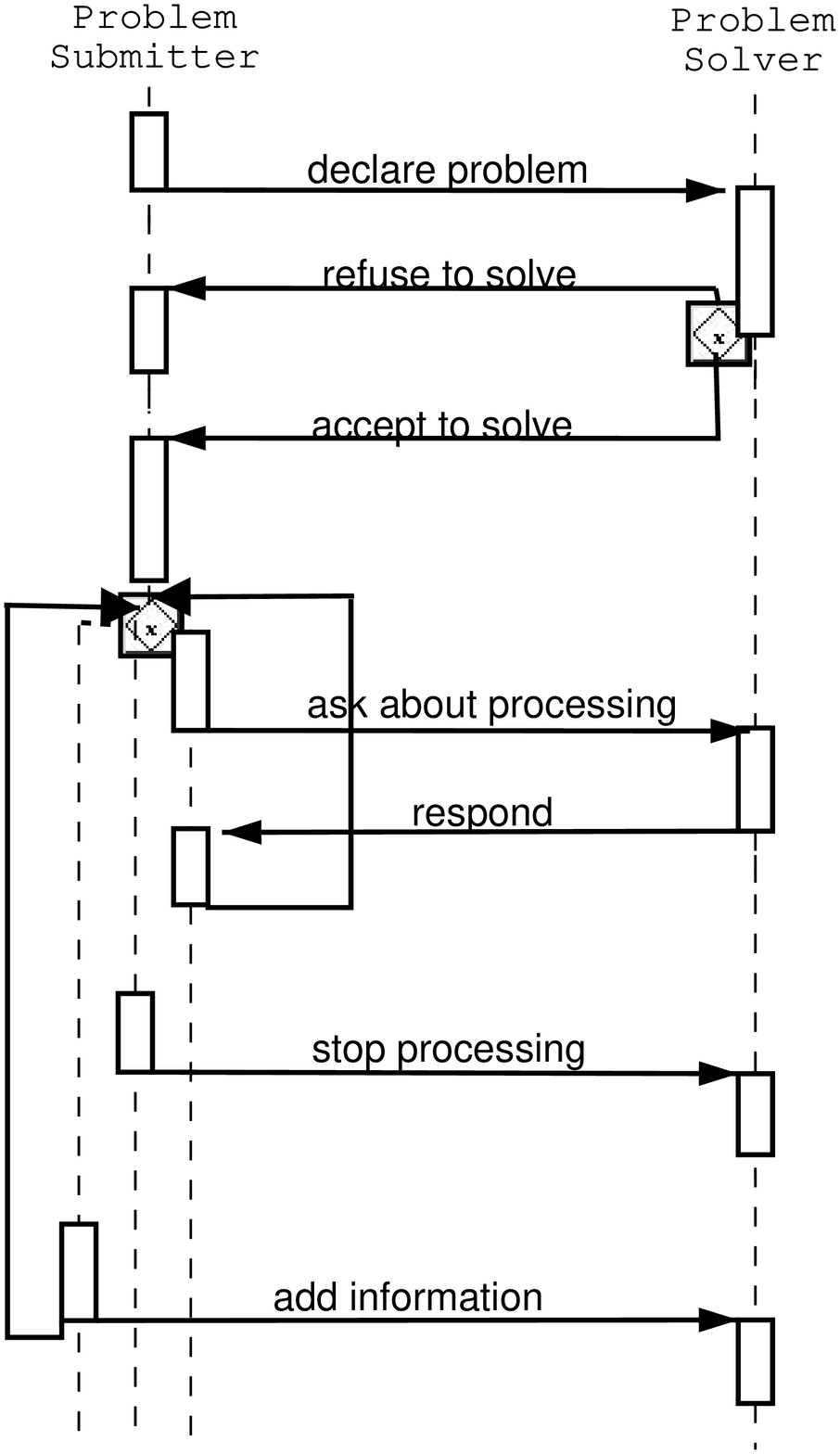}
\caption{Incremental Problem Solving Protocol}
\label{fig:incpbsolv}
\end{figure}

Consider, by way of illustration, a \emph{query agent} {\tt q$_1$}
which is requested to execute a task $t_1$: ``\emph{find out a
  document exhibiting the following characteristics:
  (language='English', content='plain/text')}''.
Protocols and potential participant agents identifications for $t_1$
lead to the matrix given in table~\ref{tab:matrix11} where a cross in
a cell [l,c] indicates that the agent at column $c$ can play a
participant role in the protocol at line $l$. This matrix reveals that
{\tt q$_1$} has identified IPS (an \emph{Incremental Problem Solving}
protocol where a problem submitter -initiator- and its solver -the
participant- progressively find out a solution to a given problem) and
Request. A diagrammatic representation of IPS is given in
figure~\ref{fig:incpbsolv}. In addition, {\tt q$_1$} identified seven
potential participant \emph{document agents} {\tt d$_1$}~\ldots~{\tt
  d$_7$}.
\begin{table}[hhhh!]
\begin{center}
\begin{tabular}{|l|c|c|c|c|c|c|c|}
\hline
& $d_1$ & $d_2$ & $d_3$  & $d_4$ & $d_5$ & $d_6$ & $d_7$\\
\hline
IPS & x & x  &  & x & x &  & x\\
\hline
Request &  &  & x & x & x &  & x\\
\hline
\end{tabular}
\end{center}
\caption{matrix for task $t_1$}
\label{tab:matrix11}
\end{table}
{\tt q$_1$} adopts a protocol-oriented exploration and selects $IPS$
(the least sparse vector). Therefore, it will try to validate $IPS$
with {\tt d$_1$}, {\tt d$_2$}, {\tt d$_4$}, {\tt d$_5$} or {\tt
  d$_7$}.
If {\tt q$_1$} receives a {\tt ready-to-select} in reply to a prior
{\tt call-for-collaboration}, it explores the list of preferred roles
and as soon as it finds the participant role of IPS or Request it
notifies its agreement with a {\tt notify-assignment}. If {\tt q$_1$}
identified more than these protocols and a role of any of them is
pointed out in the participant's {\tt ready-to-select}, this protocol
will be adopted. In the case it didn't find out any role it expects or
it received an {\tt unable-to-select}, {\tt q$_1$} replies with a {\tt
  stop-selection} and as long as there are still unexplored potential
participant agents for IPS, {\tt q$_1$} will continue contacting them.
In absence of solution when the potential participants' set has been
thoroughly explored for a protocol, the same process is taken again
upon another protocol if there is any. In case no solution has been
found and no more protocol and participant can be explored, the
dynamic protocol selection fails and the subsequent task remains not
executed.

\subsubsection{\emph{1-1$^N$} Protocols}

A solution to the dynamic \emph{1-1$^N$} Protocols selection problem
is a couple $(\mathcal{A}, p_j)$ where $\mathcal{A}$ is the set of
participant agents and $p_j$ the protocol to use. For this category of
protocols the matrix is explored only in a protocol-oriented way since
all the identified agents for a protocol are contacted at the same
time. Once all the contacted agents have replied, the initiator agent
should select a common protocol for the agents (generally for a part
of them) which replied a {\tt ready-to-select}.  We devised several
strategies to perform this selection but here we only describe one,
\emph{the largest set's strategy}, 
which looks for the role most agents selected.  If there exists a role
$r_i$ that all the agents pointed out, then this one is selected for
all the agents which replied a {\tt ready-to-select}.  Otherwise, we
look for a role that will involve the largest set of agents.
Therefore, we construct an array where each index $\imath$ contains a
collection of all the roles which exactly $\imath$ agents candidated
for. As we didn't succeed in finding out a role for all the $n$ agents
which sent a {\tt ready-to-select}, the highest index in this array
points to a collection of roles that exactly $n-1$ agents candidated
for.
We traverse the array from the highest index down to the lowest.
While exploring index $k$, if the collection of roles is not empty, we
represent for each role $r_\imath$ the set $e_\imath$ of agents which
candidated for it.

  \begin{enumerate}
    
  \item If all the $e_i$ are equal, then we randomly select a $r_p$
    and adopts its corresponding $e_p$ as the participant agents' set.
    The solution is then $e_p$ and the relevant protocol $r_p$ belongs
    to.

  \item Otherwise:

    \begin{enumerate}
      
    \item For each $e_\imath$, we compute the difference with the
      union of all the sets except $e_\imath$: ${\displaystyle
        dif_{\imath} = (e_\imath - \bigcup_\jmath\{e_\jmath\}, j \neq
        \imath)}$. Then, we select the largest $dif_\imath$.
      
    \item If no decision could be made, we save all the $e_\imath$
      (only for the highest index) and proceed on our iteration.
      
    \end{enumerate}
      
  \end{enumerate}
  
  After a solution has been found for an index $k$, we check whether
  some sets have not been saved for a higher index. If no sets were
  found the final solution is the one at hand. Otherwise, we select
  from the latters the set whose intersection with the currently
  selected set is the largest. If no solution has been found we
  iterate through the selection process changing protocols.

%

\subsubsection{\emph{1-N} Protocols}

A solution to the dynamic \emph{1-N} Protocols selection problem is a
triple $(\mathcal{A}, p_j, m)$ where $\mathcal{A}$ is the set of
participant agents, $p_j$ the protocol to use and $m$ an associative
array mapping each agent to the role(s) it will play in the protocol.
Here again, the matrix is explored only in a protocol-oriented way.
The initiator agent $a_\imath$ waits for all the participants' replies
and gathers the {\tt ready-to-select} messages. The roles are
clustered following the protocols they belong to and the protocols
which have not been identified by the initiator agent are eliminated.
For each protocol $p$, $a_\imath$ maps each role $r_\jmath$ to a set
of agents which candidated for it: ${\displaystyle
  candidates(r_\jmath)=\bigcup_k\{a_k\}}$. If ${\displaystyle
  candidates(r_\jmath)=\emptyset}$, the protocol $r_\jmath$ belongs to
is no more considered in the selection process.
Moreover, as there exists several participant roles in \emph{1-N}
protocols, some of them may receive their first message from other
participant roles. Thus, we introduce a new relation, \emph{father}:
given two roles r$_1$ and r$_2$ of a \emph{1-N} protocol,
${\displaystyle r_1 = father(r_2) \models
  r_1~is~the~sender~of~the~r_2's~first~message}$. For each protocol
retained after the candidates sets construction, the initiator agent
constructs a tree $t$ wherein nodes are the roles of the protocol. A
node r$_m$ is child of another node r$_n$ if r$_n$ = father(r$_m$).
$t$ is traversed in a breadth-first way and for each node $r_\jmath$
of $t$ an agent $a_\jmath$ is assigned to $r_\jmath$ from
${\displaystyle candidates(r_\jmath)}$. Assigning a role to an agent
can be performed by any well known resource allocation algorithm (ex
election). This assignment is achieved for all the trees and the
initiator agent uses a strategy to select one of the totally assigned
trees. An improvement during the roles assignment is to avoid
situations where the same agent plays several roles in a protocol.
Thus, when {\tt candidates} is a singleton, its only one agent is
removed from all other {\tt candidates} sets it appears in when these
are not singletons. As well, while exploring $t$, once a role has been
assigned to an agent we should remove this agent from all the {\tt
  candidates} sets it appears in provided these are not singletons.
The singleton criterion may guide a tree selection strategy.


\subsection{Beyond the joint protocol selection}

The joint protocol selection mechanism does not apply to all
interaction contexts.
One evident issue is that generic protocols are thought to be
invariably specified in the MAS. However, the only one aspect that
remains invariable in generic protocols' specification is the
description of exchanged messages which is imposed by communication
languages (KQML, FIPA ACL) and embedded in the protocols'
specification.  Hence, there is no guarantee for generic protocols to
be specified in a unique formalism and agents might fail to interpret
some of the formalisms used to specify generic protocols in the MAS.
Particularly, plugging heterogeneous sub-systems together in a MAS
increases the risk for multiple generic protocols specification
formalisms. The joint protocol selection then falls short in a MAS
where generic protocols are specified in several formalisms and agents
are unable to interpret all those formalisms.
In addition, there are also situations where agents do not always
trust one another. Then, basing protocols selection on specifications
exchange becomes unsafe.

To address these drawbacks, we developed an \emph{individual protocol
  selection} method.

\section{Individual Protocol Selection}
\label{sect:dynseli}

This selection form is carried out concomitantly to the targeted
interaction. Likewise the joint protocol selection form, the initiator
agent is in charge of starting the selection process when it locates a
collaborative task's descriptions. It finds out some protocols which
comply with the task's descriptions and wherein it can play an
initiator role.  The initiator agent may adopt a static behaviour
during this selection by choosing a protocol among the candidates. in
this case, the strategy it adopts is required to be fair. It may also
be given the possibility to exhibit dynamic behaviours by changing
protocols in order to address occurring inconsistencies. In this paper
we consider the first case.

The initiator agent sends the initial message $m_0$ of the selected
protocol $p_\imath$ to one or several potential participant agents.
$m_0$ actually denotes a need for a new interaction and any agent
which receives it selects a participant role $r_\jmath$ which starts
with $m_0$'s reception. Hence, each participant agent $a_\jmath$
constructs the \emph{collection of candidate roles} $r_\jmath$ which
we refer to in the remainder of this section as ${\displaystyle
  collection(a_\jmath, t_k)}$. The roles are then selected from
${\displaystyle collection(a_\jmath, t_k)}$ and instantiated so that
the interaction can take place. The individual protocol selection,
although more sophisticated and powerful, can lead to interaction
inconsistencies. Indeed, as individually selected roles may mismatch,
the exchanged messages' content or structure (performative, ontology,
language, etc.) may be wrong. Thus, we provide agents with techniques
to anticipate such errors by checking incoming messages over structure
and content compliance. When ${\displaystyle collection(a_\jmath,
  t_k)}$ is a singleton, the only one role is instantiated in order to
interact. If any error occurs during the interaction no recovery would
have been possible.  Dynamically selecting protocols is more appealing
when the ${\displaystyle collection(a_\jmath, t_k)}$ contains several
roles.  In this case we explore ${\displaystyle collection(a_\jmath,
  t_k)}$ either \emph{sequentially} or \emph{in parallel}. In this
paper, we only describe individual \emph{1-1 protocols} selection
since the selection mechanism is quite similar for the other two types
of protocols and only some extensions are required to fit the
specificity of these protocols.





\subsection{Sequential Roles Instantiation}


In the purpose of starting an interaction or replacing a failing role
during an interaction, a participant agent randomly (or using another
strategy we'll define later) selects roles from the collection one
after the other until there is no available role to select or the
interaction eventually safely ends up.  Once selected, roles are
removed from the collection in order to avoid selecting them anew
during the same interaction. When a message is wrong the participant
agent must recover from this error by replacing the failing role. The
recovery process lies on the interaction's journal where agents log
the executed methods and the related events (input: events which fired
the method, and output: events generated by the method's execution).
Each method and its events form a record. The following four steps
define the error recovery process:
  \begin{enumerate}
  \item If an agent detects an error, it notifies its interlocutor;
  \item $a_\jmath$ then purges ${\displaystyle collection(a_\jmath,
      t_k)}$ and selects another role;
  \item $a_\jmath$ computes the point where the interaction should
    continue at in the newly selected role and notifies the initiator;
  \item Both agents update their journals by erasing the wrong records
    and the interaction proceeds.
  \end{enumerate}
  The participant role replacement during error recovery can require
  the initiator agent to roll some actions back in order to
  synchronise with the newly instantiated role. To purge
  ${\displaystyle collection(a_\jmath, t_k)}$, $a_\jmath$:

\begin{enumerate}
  
\item Removes from ${\displaystyle collection(a_\jmath, t_k)}$ all the
  roles whose description, from the beginning of the role to the point
  the error occurs at, does not match the journal;
  
\item If the message structure is wrong and the error has been
  detected by the initiator agent, removes from ${\displaystyle
    collection(a_\jmath, t_k)}$ the roles that generated the wrong
  message. If the error has been detected by $a_\jmath$ itself, it
  removes from ${\displaystyle collection(a_\jmath, t_k)}$ the roles
  that can't receive the claimed erroneous message;
  
\item If the message content is wrong and if the error was detected by
  the initiator agent, removes from ${\displaystyle
    collection(a_\jmath, t_k)}$ the roles that cannot generate the
  same message structure at the point the error occurs at and removes
  from ${\displaystyle collection(a_\jmath, t_k)}$ the roles that use
  the same method as the one causing the error.  If the error was
  detected by the participant agent, removes from ${\displaystyle
    collection(a_\jmath, t_k)}$ the roles that do not receive the same
  message structure at the point the error occurred at and also
  removes from ${\displaystyle collection(a_\jmath, t_k)}$ the roles
  that do not receive the same message content at the point the error
  occurred at

\end{enumerate}

Since it's no use checking the content when the message structure is
wrong, structure compliance is checked prior to content's. When a role
does not comply with the current execution, it is removed from
${\displaystyle collection(a_\jmath, t_k)}$. Roles removal actually
consists in marking them so that they can no more be instantiated in
the current interaction. Then, from the updated ${\displaystyle
  collection(a_\jmath, t_k)}$, $a_\jmath$ selects a new role following
the strategy described hereby:

\begin{enumerate}
  
\item If the message content is wrong: (a) for each role of
  ${\displaystyle collection(a_\jmath, t_k)}$, construct the set of
  messages (generated or received) at the point the error occurred at;
  (b) withdraw from these sets the message that caused the error; (c)
  compare these sub sets and select the \emph{weakest} one; Then the
  role the selected sub set originates from is instantiated. The
  \emph{weakest} messages sub set is the one containing the higher
  number of weak messages. Weak messages are those which lead to
  interaction termination; these messages are potentially weaker than
  those which continue the interaction. The reason why we prefer the
  weakest messages sub set is that we wish to avoid producing another
  message than the ``structurally'' correct one we generated priorly.
  When there are several sub sets candidate for selection or when
  there are none, a role is randomly selected.
  
\item If the message structure is wrong: randomly select a role in the
  ${\displaystyle collection(a_\jmath, t_k)}$.

\end{enumerate}

Once a new role has been selected, the participant agent might expect
to continue its execution from the point the error occurred at.
However, doing so can bring inconsistencies in the interaction
execution because the roles, though enacted by the same agent, do not
necessarily use the same methods.  These inconsistencies could be
avoided by looking for methods of the new role which follow the same
sequence order as in the journal from the starting point.  This set of
methods won't be re-executed.  
We represent the methods of the new role as nodes of a directed graph
wherein an edge m$_i$m$_j$ means that method m$_j$ can be executed
immediately after m$_i$ completes and the conditions for its execution
hold.  Algorithm \ref{algo:calpra} achieves this computation and
returns the recovery points for both the initiator and the
participant.
 \begin{algorithm}[hhhh]
    \caption{recovery points computation}
    \label{algo:calpra}
    \begin{algorithmic}
      \scriptsize \STATE[Input: Journal $\equiv$ the participant's
      journal] \STATE[Input: Graph $\equiv$ the new role's methods
      graph] \STATE[Result: Initiator and participant recovery points]
      \STATE stop $\leftarrow$ false; \STATE $\imath$ $\leftarrow$ 1;
      \STATE $\jmath$ $\leftarrow$ 1; \STATE mth$\_$journal
      $\leftarrow$ read$\_$init(journal); \STATE mth$\_$graph $\leftarrow$
      get$\_$init$\_$node(graph); 
      \IF{mth$\_$journal $\neq$ mth$\_$graph} \STATE stop $\leftarrow$
      true; \ENDIF \WHILE{stop = false} \STATE mth$\_$journal
      $\leftarrow$ Succ(mth$\_$journal); 
      \IF{mth$\_$journal = null} \STATE stop $\leftarrow$ true;
      \ELSIF{mth$\_$journal $\in$ follow(mth$\_$graph)}
      \STATE mth$\_$graph $\leftarrow$ mth$\_$journal; \STATE $\jmath$
      $\leftarrow$ $\jmath$+1; \IF{Input is Message} \STATE $\imath$
      $\leftarrow$ $\imath$+1; \ENDIF \ELSE \STATE stop $\leftarrow$
      true; \ENDIF \ENDWHILE \STATE initiator recovery point:
      $\imath$; \STATE participant recovery point: $\jmath$;
    \end{algorithmic}
  \end{algorithm}
  In this algorithm, $\imath$ is the number of the latest message the
  initiator should be considering it has sent to the participant and
  $\jmath$ the point where the participant is to start executing its
  new role from. The third event type considered in the journal (data
  value change in addition to message emission and reception) accounts
  for the difference between $\imath$ and $\jmath$. Once the initiator
  receives $\imath$, it looks for the record in its journal
  representing the $\imath^{th}$ message it sent to the participant.
  All the records following this one will be erased from the journal.
  The participant also updates its journal in quite the same way
  basing on $\jmath$.
  
\begin{figure}[hhhh]
      \centering \includegraphics[width=.4\textwidth]{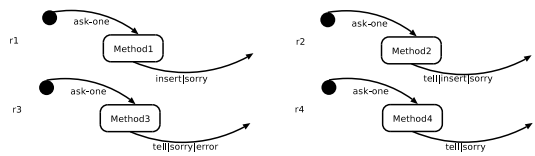}
      \caption{agent c's interaction model}
      \label{fig:partauto}
    \end{figure}

  Suppose {\tt d$_1$} wants to identify the language its document is
  written in; this task ($t_2$) requires an interaction with a
  \emph{rule agent}. We assume {\tt d$_1$} finds out a protocol which
  starts with an {\tt ask-one} emission and expects a tell. Consider
  {\tt d$_1$} contacted a \emph{rule agent} {\tt c$_1$} whose
  interaction model is partially depicted in
  figure~\ref{fig:partauto}. In this figure, for example the given
  portion of {\tt r$_1$} can be interpreted as: {\tt r$_1$} receives
  an {\tt ask-one} and can reply an {\tt insert} or a {\tt sorry}.
  ${\displaystyle collection({\tt c_1}, t_2)= \{{\tt r_1}, {\tt r_2},
    {\tt r_3}, {\tt r_4}\}}$.
  \begin{itemize}
    
  \item Whatever role {\tt c$_1$} selects, if it replies a
    \emph{sorry} the interaction will end up, may be prematurely -the
    first message has not been validated yet. To get sure it's not so
    {\tt d$_1$} issues a warning: \emph{``May be premature interaction
      termination!''}.  {\tt c$_1$} tries to select another role and
    the interaction proceeds or definitely stops.
    
  \item If {\tt c$_1$} selected {\tt r$_4$} and sent a \emph{tell}
    whose content is wrong {\tt d$_1$} notifies {\tt c$_1$}a
    \emph{wrong message content} error. {\tt c$_1$} then stops {\tt
      r$_4$}, purges its ${\displaystyle collection(a_\jmath, t_k)}$
    by removing {\tt r$_1$} (since it cannot generate a {\tt tell} at
    the error location), selects {\tt r$_3$} because it corresponds to
    the weakest sub set and computes the recovery points: $\imath =
    \jmath = Rec\sharp1$. {\tt c$_1$} updates its journal and replies
    anew to the {\tt ask-one}.

  \end{itemize}
  
  In order to avoid inconsistencies at the end of interactions, we
  require both agents to explicitly notify each other the protocol's
  termination.  Instead of selecting roles on the basis of messages
  they can generate, it sounds to select them on the basis of messages
  they really generated. This is possible only if all candidates roles
  are instantiated at the same time. This parallel roles instantiation
  is only known from the participant agent; the initiator agent still
  has the perception of a sequential roles instantiation.
  
  
\subsection{Mixed Roles Instantiation}

All the roles $a_\jmath$ identified in ${\displaystyle
  collection(a_\jmath, t_k)}$ are instantiated at the same time. They
handle the received message and generate their reply messages which
are stored in a \emph{control zone} ($C_z$); $C_z$ also contain the
messages which are destined to currently activated roles.  The roles
are then deactivated. Only one message $m_{k}$ is selected from $C_z$
and sent to the initiator. This selection can be performed following
several strategies. For example, the participant agent can randomly
select a message among those which don't shorten the interaction.
Therefore, if an {\tt insert}, a {\tt sorry}, an {\tt error} and a
{\tt tell} are generated in reply to an {\tt ask-one}, {\tt insert}
and {\tt tell} will be preferred to {\tt sorry} and {\tt error}, and
the random selection will be performed between the first two messages.
After a message $m_{k}$ has been selected, all the roles which
generated a message of the same structure and content are activated.
In this instantiation mode, when an error occurs, the participant
agent recovers from it by stopping the wrong roles and by reactivating
one or several other roles. Thus, if an error occurs on $m_{k}$:

\begin{enumerate}

\item All the activated roles are stopped;
  
\item If the message structure is wrong: all the $m_{k}$ as well as
  the messages having the same structure are removed from $C_z$ and
  their roles are stopped. The participant selects another message
  $m_{k'}$ following the same principle as $m_{k}$'s selection.
  
\item If the message content is wrong: all the $m_{k}$ messages as
  well as those having the same content pattern are removed from $C_z$
  and their roles are stopped. The participant selects another message
  $m_{k'}$ having the same structure but a different content pattern.

\end{enumerate}


When some roles stayed activated, they all generate their messages.
If the messages have the same structure and content, all these roles
remain activated. Otherwise, only one message is selected and all the
roles whose message have not been selected are deactivated. If all the
previously activated roles have been stopped, the participant agent
reactivates the most recently deactivated role but cares about early
interaction termination. When there are more than one such roles, they
all are reactivated. The participant role reactivation, might require
the initiator to roll some actions back to a recovery point.  An
algorithm similar to algorithm~\ref{algo:calpra} performs the recovery
points computation for both the initiator and participant. For each
role the algorithm is applied, considering the roles' current
execution, and the final recovery point is the earliest.

\section{Related Work}
\label{sec:related}

Protocols selection in agents interactions design is something
generally done at design time. Indeed, most of the agent-oriented
design methodologies (\emph{Gaia}~\cite{Wooldridge:00} and
\emph{MaSE}~\cite{Deloach:00} to quote a few) all make designers
decide which role agents should play for each single interaction.
However dynamic behaviours and openness in MAS demand greater
flexibility.

To date, there have been some efforts to overcome this limitation.
\cite{Durfee:99} introduces more flexibility in agents' coordination
but it only applies to planning mechanisms of the individual agents.
\cite{Boutilier:99} also proposes a framework based on multi-agent
Markov decision processes. Rather than identifying a coordination
mechanism which suits best for a situation, this work deals with
optimal reasoning within the context of a given coordination
mechanism. \cite{Bourne:00} proposed a framework that enables
autonomous agents to dynamically select the mechanism they employ in
order to coordinate their inter-related activities. Using this
framework, agents select their coordination mechanisms reasoning about
the rewards they can obtain from collaborative tasks execution as well
as the probability for these tasks to succeed.


The main requirement the selection process faces in protocol based
coordination mechanisms is whether or not there exists in the agent's
interaction model roles capable of supporting the desired interaction.
To fill this void, we proposed a method to enable agents to
dynamically select protocols basing on their interaction capacities.

\section{Conclusion}
\label{sec:conclusion}

Designing agents for open and dynamic environments is still a
challenging task, especially in regard to protocol based interactions.
Two main concerns arise from interactions modelling and design in such
systems. First, how interactions which are based on generic protocols
are configured so that consistent messages exchange can take place?
Second, does it sound that designers always decide which protocols and
roles to use every time an interaction is asked for?  We address both
issues by developing several methods. In this paper we focus on the
second concern. We argued that due to openness and dynamic behaviours
more flexibility is needed in protocols selection. Furthermore, in the
context of complex applications demanding multi-protocols agents,
moving from static to dynamic protocol selection greatly increases
such systems' efficiency and properly handles the situation tightly
related to openness where all the protocols are not known at design
time. Thus, we enabled agents to dynamically select protocols upon the
prevailing circumstances.


One outcome of the dynamic protocol selection is that the protocols to
use are no more hard-coded in all agents source code. Rather,
programmers mention collaborative tasks descriptions in the
\emph{initiator agent}'s source code only making the latter in charge
of firing the interaction.  Agents are given two ways to select
protocols.  First, the initiator agent and all (or a part of them) the
potential participant agents it identified can join together and share
information and preferences about the protocols at hand in order to
select a protocol and assign a role to each agent. Second, agents are
given the possibility to individually select their protocols and roles
anticipating errors. We focus on two types of errors: \emph{wrong
  message structure} and \emph{wrong message content}.  As roles
replacement are performed as soon as an anomaly is detected, we
constrain actions executed during interactions to be reversible and
not to render critical side effect. Furthermore, when there are
several candidate protocols in the individual protocol selection, we
developed two exploration mechanisms for these candidates: (1) a
sequential exploration and (2) a mixed exploration modes.

Both methods have been proposed and tested in the context of a
European project dedicated for information filtering. They proved
their usefulness to efficiently manage the multiple interactions that
take place between agents. In this paper, we don't provide the results
we obtained from the application of these methods since they need to
be interpreted and compared to static selection cases. In the bargain,
our aim was to describe the theoretical basis of a dynamic protocols
selection method. Our method intensively benefits from the agents'
capacity to interpret, relate and update models embedded inside them.

\bibliographystyle{plain}
\bibliography{aose}

\begin{thebibliography}{1}

\bibitem{Bourne:00}
R.~Bourne, C.~B. Excelente-Toledo, and N.~R. Jennings.
\newblock Run-time selection of coordination mechanisms in multi-agent systems.
\newblock In {\em Proceedings of the 14th European Conference on Artificial
  Intelligence}, pages 348--352, Berlin, Germany, August 2000.

\bibitem{Boutilier:99}
C.~Boutilier.
\newblock Sequential optimality and coordination in multiagent systems.
\newblock In {\em Proceedins of the Sixteenth International Joint Conference on
  Artificial Intelligence (IJCAI-99)}, pages 478--485, 1999.

\bibitem{Deloach:00}
S.~Deloach and M.~Wood.
\newblock An overview of the multiagent systems engineering methodology.
\newblock In P.~Ciancarini and M.~Wooldridge, editors, {\em Proceedings of the
  1$^{st}$ Interational Workshop on Agent Oriented Software Engineering},
  volume 1957. Springer Verlag, June 2000.

\bibitem{Durfee:99}
E.~H. Durfee.
\newblock Practically coordinating.
\newblock {\em AI Magazine}, 20(1):99--116, 1999.

\bibitem{FIPA:01b}
FIPA.
\newblock Fipa interaction protocol library specification.
\newblock Technical report, Foundation for Intelligent Physical Agents, 2001.

\bibitem{Smith:80}
Smith~R. G.
\newblock The contract net protocol: High-level communication and control in a
  distributed problem solver.
\newblock {\em IEEE Trans. On Computers}, 29(12):1104--1113, 1980.

\bibitem{Wooldridge:00}
M.~Wooldridge, N.~Jennings, and D.~Kinny.
\newblock The gaia methodology for agent-oriented analysis and design.
\newblock {\em Autonomous Agents and Multi-Agent Systems}, 3:285--312, 2000.

\end{thebibliography}

\end{document}